\documentclass[12pt]{extarticle}

\pdfoutput=1

\usepackage{amsmath}
\usepackage{amssymb}
\usepackage{calrsfs}
\usepackage{cmupint}
\usepackage{colortbl}
\usepackage{graphicx}
\usepackage{hyperref}
\usepackage{latexsym}
\usepackage{mathtools}
\usepackage{multirow}
\usepackage{rotating}
\usepackage{slashed}
\usepackage{soul}
\usepackage{xcolor}
\usepackage[normalem]{ulem}
\usepackage{cite}

\usepackage[left=2.25cm, right=2.25cm, top=2.5cm, bottom=2.5cm]{geometry}

\newcommand{\ii}{\ensuremath{\mathrm{i}}}
\newcommand{\dd}{\ensuremath{\mathrm{d}}}
\newcommand{\SU}[1]{\mathrm{SU}(#1)}

\definecolor{tbdpurple}{rgb}{0.55,0,0.65}
\newcommand{\checked}[1]{#1}

\begin{document}

\begin{center}
{\Large The pressure of the gluon plasma at $\mathcal{O}(g_s^6)$, perturbatively}
\vspace{0.8cm}

\textbf{
Mikael Chala~\footnote{mikael.chala@ugr.es}, Luis Gil~\footnote{lgil@ugr.es} and Javier López Miras~\footnote{jlmiras@ugr.es}
}\\
\vspace{1.cm}
{\em {Departamento de F\'isica Te\'orica y del Cosmos, Universidad de Granada, Campus de Fuentenueva, E--18071 Granada, Spain}}\\[0.3cm]
\vspace{0.5cm}
\end{center}

\date{}

\begin{abstract}
    We determine the four-loop hard-scale matching correction to the pressure of the gluon plasma at high temperature. This result completes the long-standing calculation through $\mathcal{O}(g_s^6)$, where $g_s$ is the strong coupling. Our result agrees, within uncertainties, with the value obtained in \cite{Borsanyi:2012ve} by fitting the perturbative series to continuum-extrapolated lattice data, thereby providing the first direct test of the weak-coupling expansion at this order.
\end{abstract}

\newpage

\section{Introduction}
Accurately determining the pressure ($p_\mathrm{QCD}$) of the gluon plasma at high temperature ($T$) remains a major theoretical challenge. It is essential for predicting the production and decoupling of cosmological relics, including light sterile neutrinos~\cite{Asaka:2006rw} whose production can peak at $T\sim\Lambda_{\mathrm{QCD}}$; modeling the hydrodynamic evolution of matter created in heavy-ion collisions and the resulting hadron momentum spectra~\cite{Huovinen:2009yb,Shen:2010uy}; characterizing the strength of first-order phase transitions~\cite{Espinosa:2010hh}; or predicting the imprint of the QCD transition on the spectrum of primordial gravitational waves generated during inflation~\cite{Schwarz:1997gv}.

At high $T$, the full pressure $p_\mathrm{QCD}$ admits an expansion in the strong coupling $g_s$, including logarithmic terms, %of the form $p(T)=\sum_n c_n(T)g_s^n$, 
where the leading order $8\pi^2 T^4/45$ is the Stefan--Boltzmann law, whereas the $\mathcal{O}(g_s^2)$ and $\mathcal{O}(g_s^3)$ corrections date back to the late 70's~\cite{Shuryak:1977ut,Chin:1978gj,Kapusta:1979fh}, the $\mathcal{O}(g_s^4\log g_s)$ term to the 80's~\cite{Toimela:1982hv}, and the $\mathcal{O}(g_s^4)$ and $\mathcal{O}(g_s^5)$ ones to the 90's~\cite{Arnold:1994ps,Arnold:1994eb,Zhai:1995ac,Braaten:1995jr}. Partial $\mathcal{O}(g_s^6)$ results are also known~\cite{Kajantie:2002wa,DiRenzo:2006nh,Gynther:2009qf,Navarrete:2024ruu}.
The modern approach to the problem actually splits the calculation in three pieces: $p_\mathrm{QCD}=p_\text{hard}+p_\text{soft}+p_\text{ultrasoft}$. They capture the contributions from the scales $\sim \pi T$, $\sim g_s T$ and $\sim g_s^2 T$, respectively~\cite{Braaten:1995jr}, in a process known as \textit{dimensional reduction}~\cite{Ginsparg:1980ef,Appelquist:1981vg,Braaten:1995cm,Kajantie:1995dw}.
These can be in turn seen as different contributions coming from each layer in a tower of 3-dimensional (3D) effective field theories (EFT).
The first one, known as \textit{electrostatic} EFT, keeps dynamical only the zeroth modes of fields that have been Fourier-expanded around the compactified temporal direction
of radius $(2 \pi T)^{-1}$~\cite{Matsubara:1955ws} while all other (Matsubara) modes are integrated out. The second \textit{magnetostatic} EFT results from further integrating out the adjoint scalar (the zeroth mode of the gluon), which acquires a Debye screening mass~\cite{Gross:1980br,Braaten:1995jr,Ghisoiu:2015uza,Chala:2026wjp} in the first step of dimensional reduction.

Thus, $p_\text{hard}$ is the contribution associated with the matched coefficient of the electrostatic unit operator; $p_\text{soft}$ contains the perturbative vacuum contributions obtained by integrating out the massive electric modes within this EFT; and $p_\text{ultrasoft}$ is the vacuum pressure of the magnetostatic EFT. To reach 
$\mathcal{O}(g_s^6)$ accuracy, $p_\text{hard}$ requires four-loop matching, while $p_\text{soft}$ involves vacuum diagrams through four EFT loops~\cite{Kajantie:2003ax}, with the 3D mass parameter matched through two loops and the gauge and scalar quartic couplings through one loop~\cite{Kajantie:2002wa}. The ultrasoft contribution first enters at $\mathcal{O}(g_s^6)$. Its logarithmic scale dependence is accessible perturbatively at four loops, whereas its finite constant is intrinsically nonperturbative~\cite{Kajantie:2002wa,DiRenzo:2006nh}. Out of all these contributions, the only remaining ingredient at this order is the finite four-loop hard matching coefficient.

Major progress was recently made in \cite{Navarrete:2024ruu} in this respect, where the pure four-loop contribution to $p_\text{hard}$ was reduced to the evaluation of 10 non-factorizable four-loop sum-integrals. In this paper, we carry out an independent calculation, using a different basis of sum-integral
propagators, and provide the $\epsilon$-expansion of all required sum-integrals to finite order, in dimensional regularization in $d=3-2\epsilon$ dimensions. We thus complete the $\mathcal{O}(g_s^6)$ pressure of the hot gluon plasma, culminating more than 30 years of work on this problem.

\section{Conventions}
We consider the QCD Lagrangian in the background-field gauge:
\begin{align}
    \mathcal{L} &= -\frac{1}{4} G_{\mu\nu}^A G^{A\mu\nu}-\overline{c}^A \bigl(\overline{D}^\mu D_{\mu}c\bigr)^A-\frac{1}{2\xi} \bigl(\overline{D}^\mu Q_{\mu}\bigr)^A\bigl(\overline{D}^\nu Q_{\nu}\bigr)^A
\end{align}
$G$, $Q$ and $c$ represent the background, quantum and ghost fields, respectively; and we take $\xi=1$. Our convention for the covariant derivative is
\begin{align}
    (D_\mu)^{AB} &= \delta^{AB}\partial_\mu+ g_s f^{ACB}G_\mu^C\,.
\end{align}
$\overline{D}$ indicates that only the classical field enters the expression.

\begin{figure}[t]
    \centering
    \includegraphics[width=\textwidth]{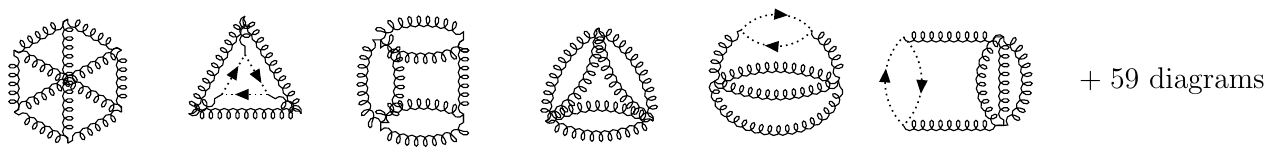}
    \caption{\it Four-loop QCD vacuum diagrams, drawn with \texttt{TikZ-Feynman}~\cite{Ellis:2016jkw}, involving gluons and ghosts. Six representative diagrams are shown out of the $65$ that contribute in the Feynman gauge, built on $26$ one-particle-irreducible topologies.}\label{fig:qcd_diagrams}
\end{figure}

Our aim is to compute the hard-scale matching of the unit operator at four loops; see Fig.~\ref{fig:qcd_diagrams}.
To this aim, we generate diagrams and vertices using \texttt{FeynArts}~\cite{Hahn:2000kx} and \texttt{FeynRules}~\cite{Alloul:2013bka}, respectively~\footnote{By default, \texttt{FeynArts} does not provide four-loop starting topologies. We construct them by closing the two external legs of the topologies generated by   \texttt{CreateTopologies[3,2,Adjacencies->\{3,4\},ExcludeTopologies->\{\},CTOrder->0]}, removing isomorphic duplicates and retaining the one-particle-irreducible graphs.}. We simplify the $\SU{N_c}$ algebra and contract the Lorentz indices with \texttt{FeynCalc}~\cite{Shtabovenko:2023idz}. All numerators in the sum-integrals described by these vacuum diagrams are polynomials in the four-dimensional scalar products of loop momenta $P_i\cdot P_j$, while no tensor reduction is required and no temporal components $P_0$ appear in numerators.

We extend \texttt{SIRENA}~\cite{Gil:2026cqz} to perform general integer unimodular changes of loop momenta beyond the default momentum shifts that only induce a permutation of the terms in the selected family of propagators. This yields linear relations among sum-integrals with different numerators, which we solve using the equation solving facilities therein, namely the ordering of the sum-integrals by complexity \cite{Laporta:2000dsw}, with a preferred set of sum-integrals prioritized as masters, Gauss forward elimination and backwards substitution. No genuine four-loop integration-by-parts (IBP) identity is used.
This way, writing $p_\text{hard}^\text{bare} = g_s^2 p_2 + g_s^4 p_3 + g_s^6 p_4$, prior to 4D renormalization, we obtain:
\begin{align}\label{eq:p2}
    p_2 &= -6(d-1)^2\,\mathcal{I}_{1}^2 \,,
\end{align}
\begin{align}\label{eq:p3}
    p_3 &= 9(d-1)^2\bigg[\mathcal{I}_{110011}+2(d-5)\mathcal{I}_{2}\mathcal{I}_1^2 +2\mathcal{I}_{21111-2}\bigg]\,,
\end{align}
as well as
\begin{align}\label{eq:p4}
    p_4=\sum_i w_i\,X_i\,,
\end{align}
where $X_i$ runs over the ten connected sum-integrals $C_i$ and the six factorizable ones $B_i$ listed in Tab.~\ref{tab:sumintegrals}, together with their coefficients $w_i$.
The subindices in the sum-integrals $\mathcal{I}$ indicate the powers of the corresponding denominators:
\begin{align}
    \label{eq:sumintdef}\mathcal{I}_{a_1\ldots a_L b_1\ldots b_n} &= \sumint_{P_1,P_2,\cdots,P_L} \frac{1}{P_1^{2a_1}\cdots P_L^{2a_L}(P_1-P_2)^{2b_1}\cdots (P_{L-1}-P_{L})^{2 b_n}}\,,
\end{align}
where $L$ counts the number of loops and $n=L(L-1)/2$, as 
we include all different pairs $(P_i-P_j)^2$. This is the standard basis of denominators in \texttt{SIRENA} and it differs from that used in \cite{Navarrete:2024ruu}. We explicitly check that both expressions are equivalent upon appropriate changes of loop variables. Note that no two-loop sum-integral arises in Eqs.~\eqref{eq:p2} to \eqref{eq:p4} because they all factorize into products of one-loop ones~\cite{Davydychev:2023jto,Gil:2026cqz}.
\begin{table}[t]
    \begin{center}
    \small
    \renewcommand{\arraystretch}{1.2}
    \begin{tabular}{llcrrr}
        \hline
        & & $w_i$ & $a$ & $b$ & $c$ \\
        \hline
        $C_1$ & $\mathcal{I}_{0011011110}$ & $-9(d^3+25d^2-293d+651)$ & \checked{$0.0624$} & \checked{$1.224$} & \checked{$13.67$} \\
        $C_2$ & $\mathcal{I}_{0011101101}$ & $324(d-1)^2$ & \checked{$0.0208$} & \checked{$0.402$} & \checked{$4.48$} \\
        $C_3$ & $\mathcal{I}_{-1011111110}$ & $648(d-9)(d-3)$ & \checked{$0.0053$} & \checked{$0.343$} & \checked{$4.45$} \\
        $C_4$ & $\mathcal{I}_{-2011011112}$ & $432(d-1)^2$ & \checked{$0.0741$} & \checked{$0.825$} & \checked{$7.643$} \\
        $C_5$ & $\mathcal{I}_{-1-111-111113}$ & $-36(d-1)^3$ & $-0.02$ & $-0.2$ & $-2$ \\
        $C_6$ & $\mathcal{I}_{-1111111-111}$ & $-162(5d-13)$ & \checked{$-0.0415$} & \checked{$-0.820$} & \checked{$-9.22$} \\
        $C_7$ & $\mathcal{I}_{-2111111011}$ & $-108(d-1)^2$ & \checked{$0.010$} & \checked{$0.668$} & \checked{$8.78$} \\
        $C_8$ & $\mathcal{I}_{-2011111111}$ & $432(d-1)^2$ & \checked{$-0.104$} & \checked{$-1.18$} & \checked{$-11.2$} \\
        $C_9$ & $\mathcal{I}_{-1-111111111}$ & $-324(d-1)^2$ & \checked{$-0.055$} & \checked{$-0.611$} & \checked{$-5.63$} \\
        $C_{10}$ & $\mathcal{I}_{-3111111111}$ & $-36(d-1)^2$ & \checked{$-0.333$} & \checked{$-4.09$} & \checked{$-42.0$} \\
        \hline
        $B_1$ & $\mathcal{I}_{-1011111011}$ & $-54(d^3+21d^2-165d+335)$ & $-0.02083$ & $-0.1674$ & $-1.2806$ \\
        $B_2$ & $\mathcal{I}_{2211000000}$ & $-\frac{54(d-1)^2(d^4-17d^3+89d^2-151d+46)}{(d-5)(d-2)}$ & $0.006944$ & $0.06044$ & $0.3281$ \\
        $B_3$ & $\mathcal{I}_{2110001010}$ & $-108(d-1)^3$ & $0.01042$ & $0.09935$ & $0.7253$ \\
        $B_4$ & $\mathcal{I}_{2211110-200}$ & $-216(d-1)^3$ & $0.03472$ & $0.2999$ & $2.0024$ \\
        $B_5$ & $\mathcal{I}_{3111110-200}$ & $-108(d-1)^3$ & $-0.01157$ & $-0.1004$ & $-0.7076$ \\
        $B_6$ & $\mathcal{I}_{3111000000}$ & $36(d-1)^3(13-d)$ & $0$ & $0$ & $0.001391$ \\
        \hline
    \end{tabular}
    \end{center}
    \caption{\it Sum-integrals entering $p_4$ in Eq.~\eqref{eq:p4}, their coefficients $w_i$, and the coefficients $a$, $b$, $c$ of their $\epsilon$-expansion in Eq.~\eqref{eq:scale}, at $\overline{\mu}=2\pi T$, i.e. $\ell=0$. For the connected $C_i$, the digits shown are significant figures from Monte Carlo errors ($\pm 1$ error on last digit). The factorizable $B_i$ are obtained from the products of Eq.~\eqref{eq:factorizable}; their numerical uncertainties are below the last digit shown.}\label{tab:sumintegrals}
\end{table}

\section{Numerical procedure}\label{sec:numerical}
Many of the sum-integrals above have been previously evaluated in the literature~\cite{Arnold:1994ps,Braaten:1995jr,Gynther:2007bw,Navarrete:2024ruu}. The one-loop ones read, in the $\overline{\rm MS}$ scheme,
\begin{align}
    \mathcal{I}_a\equiv\sumint_P\frac{1}{(P^2)^a} &= \left(\frac{e^{\gamma_E}\overline{\mu}^2}{4\pi}\right)^{\epsilon}\frac{2T\,\zeta(2a-d)\,\Gamma(a-d/2)}{(4\pi)^{d/2}\,\Gamma(a)\,(2\pi T)^{2a-d}}\,,
\end{align}
so that $\mathcal{I}_1=\frac{T^2}{12}\left[1+2\epsilon(\ell-\log2+1+Z_1)+\mathcal{O}(\epsilon^2)\right]$ and $\mathcal{I}_2=\frac{1}{(4\pi)^2}\left[\frac{1}{\epsilon}+2\ell-2\log2+2\gamma_E+\mathcal{O}(\epsilon)\right]$, where $\ell=\log{\frac{\overline\mu}{2\pi T}}$ and $Z_n=\zeta'(-n)/\zeta(-n)$, $\zeta(x)$ denoting the Riemann zeta function. The two genuine three-loop sum-integrals in $p_3$ are~\cite{Arnold:1994ps,Navarrete:2024ruu}
\begin{align}
    \mathcal{I}_{110011} &= \frac{T^4}{24(4\pi)^2}\left[\frac{1}{\epsilon}+6\ell+\frac{91}{15}-6\log2+8Z_1-2Z_3+\mathcal{O}(\epsilon)\right]\,,\\
    \mathcal{I}_{21111-2} &= \frac{T^4}{216(4\pi)^2}\left[\frac{11}{\epsilon}+66\ell+\frac{73}{2}-66\log2+12\gamma_E+64Z_1-10Z_3+\mathcal{O}(\epsilon)\right]\,.
\end{align}
The factorizable four-loop sum-integrals in $p_4$ are, exactly, products of one-loop tadpoles and three-loop sum-integrals. They read
\begin{align}\label{eq:factorizable}
    B_1 &=\mathcal{I}_{-1011111011}=\mathcal{I}_1\,\mathcal{I}_{111110}\,,\nonumber\\
    B_2 &=\mathcal{I}_{2211000000}=\mathcal{I}_1^2\,\mathcal{I}_2^2\,,\nonumber\\
    B_3 &=\mathcal{I}_{2110001010}=\mathcal{I}_1\,\mathcal{I}_{210011}\,,\nonumber\\
    B_4 &=\mathcal{I}_{2211110-200}=\frac{\mathcal{I}_1}{2}\Big[\frac{4d-22}{d-5}\,\mathcal{I}_1\mathcal{I}_2^2-\mathcal{I}_{210011}\nonumber\\&\qquad\qquad\qquad\,\,\,-2\,\mathcal{I}_{111110}-4(d-4)\,\mathcal{I}^{020}_{211110}\Big]\,,\nonumber\\
    B_5 &=\mathcal{I}_{3111110-200}=\mathcal{I}_1\,\mathcal{I}_{31111-2}\,,\nonumber\\
    B_6 &=\mathcal{I}_{3111000000}=\mathcal{I}_1^3\,\mathcal{I}_3\,,
\end{align}
where the superindices in $\mathcal{I}^{020}_{211110}$ indicate the powers of temporal components of the loop momenta in the numerator, \textit{i.e}. it is $\mathcal{I}_{211110}$ with an additional factor $(P_2^0)^2$. Since $\mathcal{I}_1$ starts at $\mathcal{O}(\epsilon^0)$, these three-loop sum-integrals are needed only through their finite parts (their $\mathcal{O}(\epsilon)$ terms do not hit any pole). We take them from the literature, converting to our $\overline{\rm MS}$ normalization. Concretely: $\mathcal{I}_{111110}$ from \cite{Schroder:2012hm}, $\mathcal{I}_{210011}$ from \cite{Gynther:2007bw}, $\mathcal{I}^{020}_{211110}$ from \cite{Ghisoiu:2012yk} and $\mathcal{I}_{31111-2}$ from \cite{Ghisoiu:2012kn}.

We evaluate the remaining ones numerically, using the method presented in \cite{Chala:2026wjp}, which is in turn based on \cite{Harnett:2024nny,Niedermayer:2016ilf,Nishimura:2018wla,Cavalcanti:2021xtb}. It can be summarized in the following steps:

\begin{enumerate}
    \item Use Schwinger parametrization to write the propagators as
    $$\frac{1}{P^{2a}} = \frac{1}{\Gamma(a)}\int_0^\infty \dd s\, s^{a-1} e^{-sP^2}\,.$$
    Then the spatial integration can be trivially carried out: writing the line momenta as $K_i=\sum_{j=1}^L C_{ij}P_j$ and $K_i^2=\vec k_i^{\,2}+\omega_i^2$, with $\omega_i=\sum_j C_{ij}\,\omega_{n_j}$ and $\omega_n=2\pi Tn$ for bosons, the $d$-dimensional integrals are Gaussian,
    \begin{align*}
    \sumint_{P_1\ldots P_L}\prod_{i}\frac{1}{(K_i^2)^{a_i}} &=\frac{T^L}{(4\pi)^{Ld/2}}\prod_i\frac{1}{\Gamma(a_i)}\int_0^\infty\dd s_i\,s_i^{a_i-1}\,\big[\det A(\mathbf{s})\big]^{-d/2}\sum_{n_1,\ldots,n_L\in\mathbb{Z}}e^{-\boldsymbol{\omega}^{T}A(\mathbf{s})\,\boldsymbol{\omega}}\,,
    \end{align*}
    where $A(\mathbf{s}) =C^{T}\mathrm{diag}(\mathbf{s})\,C$, $\boldsymbol\omega=(\omega_{n_1},\ldots,\omega_{n_L})$ and the $\overline{\rm MS}$ factor $(e^{\gamma_E}\overline\mu^2/4\pi)^{L\epsilon}$ is left implicit. Numerators ($a_i<0$) are obtained from the corresponding Gaussian moments.
    \item Split the domain of integration in \textit{sectors}, each one defined by an ordering of the rescaled $x_i\equiv (2\pi T)^2 s_i$ parameters.
    \item Split each sector into \textit{subsectors}, defined by which $x_i$ are \textit{short} and which are \textit{large}, namely which fulfill $0\leq x_i\leq 1$ and which $x_i\geq 1$, respectively. 
    \item Build the matrix $C$ of line momenta, defined by $K_i = \sum_{j=1}^L C_{ij} P_j$, where $\vec{K}$ is a subset of the denominators $\lbrace P_1, \ldots P_L,(P_1-P_2),\ldots (P_{L-1}-P_L)\rbrace$.
    \item Build the submatrix $C_\ell$ of $C_{ij}$ that involves only the large lines.
    \item Define $r\equiv \text{rank}(C_\ell)$ and $s\equiv$ number of nonzero columns in $C_\ell$.
    \item If $s>r$, perform a change of loop variables that makes the transformed large-line matrix have exactly $r$ nonzero columns. A constructive choice is $\vec{P}'=U\vec{P}$ with $U=(u^T)^{-1}$
     where $u$ is the unimodular matrix obtained from Hermite decomposition of $C_\ell$. This confines all large-line dependence to $r$ loop variables.
    \item Perform Poisson transformation of the $L-r$ transformed frequency variables that do not enter the large lines, retaining the other $r$ frequency sums, by means of the following identity:
    $$\sum_{n\in\mathbb{Z}} e^{-t (n+\delta)^2} = \sqrt{\frac{\pi}{t}}\sum_{w\in\mathbb{Z}}e^{2\pi\ii w\delta}e^{-\frac{\pi^2 w^2}{t}}\,.$$
    \item Map each such subsector to the unit cube. In a given subsector with ordering $x_{\sigma(1)}\geq x_{\sigma(2)}\geq\cdots\geq x_{\sigma(N)}$, of which the first $n_\ell$ are large, the change of variables reads
    $$x_{\sigma(j)}=\prod_{m=j}^{n_\ell}\frac{1}{t_m}\quad (j\leq n_\ell)\,,\qquad x_{\sigma(j)}=\prod_{m=n_\ell+1}^{j}t_m\quad (j>n_\ell)\,,\qquad 0\leq t_m\leq1\,.$$
    Its Jacobian is a monomial in the $t_m$: the large parameters are parameterized by inverse products of the $t_m$ and the small ones by direct products, so that every boundary of the subsector ($x_{\sigma(j)}=x_{\sigma(j+1)}$, $x_{\sigma(n_\ell)}=1$, $x_{\sigma(1)}\to\infty$, $x_{\sigma(N)}\to0$) is mapped onto a face of the cube.
\end{enumerate}

As a result, after factoring out endpoint powers, the integral in each subsector has the form
\begin{align}
    \overline{\mathcal{I}}(\epsilon) = \int_{[0,1]^N} \dd^N t \prod_{i=1}^N t_i^{d_i+c_i\epsilon} F(\mathbf{t},\epsilon)\,.
\end{align}
where $F$ is sufficiently differentiable and analytic near $\epsilon\to 0$. The overline refers to the sum-integral restricted to 
%the
a particular subsector.

Logarithmic endpoint singularities occur for $c_i\neq 0,d_i=-1$ 
%.
and these can be analytically subtracted. For example, assuming only $d_1=-1$ and denoting by $G(t_1,\epsilon)$ the integral over the remaining variables, $G(t_1,\epsilon)\equiv\int_{[0,1]^{N-1}}\prod_{i\geq2}\dd t_i\,t_i^{d_i+c_i\epsilon}F(\mathbf{t},\epsilon)$, with Taylor expansion given by $G=G_0(t_1)+\epsilon\,G_1(t_1)+\mathcal{O}(\epsilon^2)$,
\begin{align}
    \overline{\mathcal{I}}(\epsilon) &= \int_0^1\dd t_1\,t_1^{-1+c_1\epsilon}\,G(t_1,\epsilon)= \frac{G(0,\epsilon)}{c_1\epsilon}+\int_0^1\dd t_1\,t_1^{-1+c_1\epsilon}\left[G(t_1,\epsilon)-G(0,\epsilon)\right]\nonumber\\
    &=\frac{G_0(0)}{c_1\epsilon}+\frac{G_1(0)}{c_1}+\int_0^1\frac{\dd t_1}{t_1}\left[G_0(t_1)-G_0(0)\right]+\mathcal{O}(\epsilon)\,.
\end{align}
The pole is thus given by a lower-dimensional integral and the finite part by integrals with integrable kernels
%;
while higher orders in $\epsilon$ bring powers of $\log t_1$. Higher-order poles (occurring when $d_i\leq-2$, or when several $d_i=-1$) are treated in the same way: subtracting the corresponding Taylor polynomial of $G$ at the endpoint and iterating over the variables. The finite integrals can finally be evaluated numerically, upon truncating the Matsubara and Poisson sums.

For our purposes, we keep $n=4$ Matsubara modes, and use up to 16 runs  of 4096 Sobol points in the most computationally demanding sum-integrals. We estimate errors as the standard deviations among the outputs of these runs.
Exploiting the analytic relations between poles at different orders, our results are shown in Tab.~\ref{tab:sumintegrals} for $a,b,c$ in
\begin{align}
    \label{eq:scale}\mathcal{I} = \frac{T^4}{(4 \pi)^4}\bigg[\frac{a}{\epsilon^2}+\frac{b+8a\ell}{\epsilon}+c +8b\ell+32 a\ell^2+\mathcal{O}(\epsilon) \bigg]\,.
\end{align}

We use the same approach to evaluate the $\mathcal{O}(\epsilon)$ terms of the sum-integrals in $p_3$, which
hit the pole of the $g_s$ counterterm in the $\mathcal{O}(g_s^6)$ \textit{renormalized} result to yield a finite contribution. We obtain 
\begin{align}
    (4\pi)^2 T^{-4}\mathcal{I}_{110011}\big|_{\epsilon^1}&=6.636(9)\,, \\
    (4\pi)^2 T^{-4}\mathcal{I}_{21111-2}\big|_{\epsilon^1}&=4.06(5)\,,
\end{align}
at $\overline\mu=2\pi T$.

\section{Analytic evaluation of poles}\label{sec:poles}
We cross-check the previously obtained poles by requiring that, after four-dimensional UV renormalization, the left-over poles in the matching cancel those of the EFT. At four-loop order, the EFT vacuum-energy counterterm receives contributions from four-loop vacuum diagrams as well as from two-loop vacuum diagrams with two-loop scalar-mass counterterm insertions (no divergences arise in odd loop order diagrams in $d=3-2\epsilon$).
In pure hot QCD, this constrains only one combination of the ten connected sum-integrals in Tab.~\ref{tab:sumintegrals}. To obtain additional independent constraints, we perform the four-loop pressure hard-to-soft matching in the full bosonic Standard Model (SM).

This requires computing the four-loop vacuum-energy
counterterm in the corresponding 3D EFT, that we do following the massive vacuum-integral approach of \cite{Kajantie:2003ax}. We work in Feynman gauge ($\xi=1$) and assign a common positive auxiliary mass to all propagators, including those of the gauge fields and ghosts, to regulate IR divergences.

After contracting Lorentz indices in $d=3-2\epsilon$,
we map the resulting scalar integrals to the denominator families of \texttt{FMFT}~\cite{Pikelner:2017tgv}, mostly using the routines \texttt{FCLoopToPakForm} and \texttt{FCLoopFindMomentumShifts} of \texttt{FeynCalc}. 
We use \texttt{FMFT} to reduce the integrals at symbolic dimension, and insert the analytic expressions collected in Appendix~A.1 of \cite{Schroder:2025rka}. Constants without known analytic expressions are retained symbolically and cancel from the final pressure poles. The results are given in Appendix \ref{app:counterterm}.

This way, pole cancellation for all monomials, namely $g_s^6$, $g_2^6$, $g_1^2g_2^4$, $g_1^4g_2^2$, $g_1^6$, $g_2^4\lambda$, $g_1^2g_2^2\lambda$, $g_1^4\lambda$, $g_2^2\lambda^2$, $g_1^2\lambda^2$ and $\lambda^3$ (in the absence of fermions, gluons do not couple to the electroweak sector, so no mixed $g_s^4g_{1,2}^2$ or $g_s^4\lambda$ terms arise), provides $22$ linear equations, one for the double and one for the single pole of each of the $11$ monomials~\footnote{One may wonder whether extending the SM 3D EFT with further fields, and therefore more couplings, allows us to generate a larger-rank system of equations. The answer is negative, at least for extensions with up to two fields in any combination of trivial, fundamental and adjoint representations of the SM gauge groups.}, whose rank is $14$ in the $20$ pole coefficients $a_i,b_i$ of the connected $C_i$ in Tab.~\ref{tab:sumintegrals}. Equivalently, they fix the poles of $7$ independent linear combinations of sum-integrals.
Fixing
\begin{equation}
   \widehat C_i\equiv\frac{(4\pi)^4}{T^4}C_i,
   \qquad d=3-2\epsilon,\qquad \overline{\mu}=2\pi T,
\end{equation}
we obtain the pole coefficients collected in Tab.~\ref{tab:poles}, in the notation of Eq.~\eqref{eq:scale}.
\begin{table}[t]
    \begin{center}
    \small
    \renewcommand{\arraystretch}{1.6}
    \begin{tabular}{lccrr}
        \hline
        & $a$ & $b$ & $a$ & $b$ \\
        \hline
        $\widehat C_1$ & $\frac{1}{16}$ & $\frac{11}{20}-\frac{\pi^2}{32}+\frac{3Z_1}{4}-\frac{Z_3}{4}-\frac{\log2}{2}$ & $0.06250$ & $1.222$ \\
        $\widehat C_5$ & $-\frac{1}{54}$ & $-\frac{581}{4320}-\frac{\gamma_E}{18}+\frac{3\pi^2}{1024}-\frac{Z_1}{9}+\frac{Z_3}{54}+\frac{4\log2}{27}$ & $-0.01852$ & $-0.2436$ \\
        $\widehat C_{9}$ & $-\frac{1}{18}$ & $-\frac{1031}{8640}-\frac{\gamma_E}{6}-\frac{\pi^2}{128}-\frac{Z_1}{3}+\frac{Z_3}{18}+\frac{4\log2}{9}$ & $-0.05556$ & $-0.6104$ \\
        $\widehat C_2+\widehat C_3$ & $\frac{5}{192}$ & $\frac{611}{5760}-\frac{\gamma_E}{12}+\frac{23Z_1}{48}-\frac{3Z_3}{16}-\frac{5\log2}{24}$ & $0.02604$ & $0.7437$ \\
        $\widehat C_{6}-8\widehat C_2$ & $-\frac{5}{24}$ & $\frac{1}{16}-\frac{\pi^2}{12}-\frac{5Z_1}{2}+\frac{5Z_3}{6}+\frac{5\log2}{3}$ & $-0.2083$ & $-4.029$ \\
        $\widehat C_{7}+5\widehat C_2$ & $\frac{11}{96}$ & $\frac{61}{144}-\frac{\gamma_E}{6}+\frac{41Z_1}{24}-\frac{5Z_3}{8}-\frac{11\log2}{12}$ & $0.1146$ & $2.680$ \\
        $\widehat C_{10}-12\widehat C_{8}-6\widehat C_2-12\widehat C_4$ & $-\frac{7}{72}$ & $-\frac{5291}{4320}+\frac{5\gamma_E}{12}+\frac{9\pi^2}{64}-2Z_1+\frac{29Z_3}{36}+\frac{7\log2}{9}$ & $-0.09722$ & $-2.507$ \\
        \hline
    \end{tabular}
    \end{center}
    \caption{\it Double- and single-pole coefficients $a$ and $b$, defined as in Eq.~\eqref{eq:scale}, of the seven combinations of connected sum-integrals fixed by pole cancellation, at $\overline\mu=2\pi T$: exact expressions (left) and their numerical values (right).}\label{tab:poles}
\end{table}

The numerical values in Tab.~\ref{tab:sumintegrals} reproduce well these analytical results, giving further confidence of the accuracy of our numerical approach.

\section{Results}\label{sec:results}
Altogether, the renormalized hot gluon pressure through $\mathcal{O}(g_s^6)$ reads~\cite{Kajantie:2002wa}
\begin{align}
    \label{eq:strict}\frac{p}{p_{\mathrm{SB}}} = 1&-\frac{15}{4}x+30x^{3/2}+x^2\left(237.2+\frac{135}{2}\log{x}-\frac{165}{8}\ell\right)+x^{5/2}\left(-799.1+\frac{495}{2}\ell\right)\nonumber\\
    &+x^3\left[q_c(0)+\left(-1134.8+\frac{1485}{2}\ell\right)\log{x}-\frac{1815}{16}\ell^2+2932.9\ell\right]\,;
\end{align}
with
\begin{equation}
    x=\frac{g_s^2(\overline\mu)}{4\pi^2}\,,\quad p_{\mathrm{SB}}=\frac{8\pi^2T^4}{45}\,.
\end{equation}
Following the same reference, $q_c(0) \equiv q_c(N_f=0)=\frac{45}{32}\beta_{E6}(\ell=0)+q_\text{soft}+\frac{1215}{4}\beta_G$, where $\beta_{E6}(\ell)$ is the finite hard-matching $g_s^6$ term, here needed at $\ell=0$; $q_\text{soft}$ collects the known contributions from the electrostatic EFT, the magnetic logarithmic constant, and from the lower-order matching coefficients, all evaluated at $\overline\mu=2\pi T$,
\begin{align}
    q_\text{soft}=\frac{45}{32}\Big\{&-120-6\big[\alpha_{E6}+\alpha_{E5}\alpha_{E7}+3\alpha_{E7}+\beta_{E2}+\beta_{E3}\big]+24(\alpha_{E6}+\alpha_{E7})\log2\nonumber\\
    &+216\big[\beta_M-8\alpha_M\log2-8\alpha_G\log(12\pi)\big]\Big\}=-3630.97\,,
\end{align}
with the matching coefficients $\alpha_{Ei}$, $\alpha_M$ and $\alpha_G$ from \cite{Kajantie:2002wa}, the $\mathcal{O}(\epsilon)$ matching coefficients $\beta_{E2,E3}$ from \cite{Laine:2005ai}, and the four-loop vacuum constant of the electrostatic EFT $\beta_M=-1.391512$ from \cite{Kajantie:2003ax}; and $\beta_G=-0.2\pm0.4_{\rm MC}\pm0.4_{\rm NSPT}$ is the non-perturbative constant of the magnetostatic EFT, determined on the lattice in \cite{DiRenzo:2006nh}, so that $\frac{1215}{4}\beta_G=-60.8\pm121.5\pm121.5$. Finally, $\beta_{E6}$ is given by a bare four-loop contribution (that we compute in this work), and lower loops with counterterms:
\begin{align}
    \beta_{E6}(\ell)=\left[\frac{(4\pi)^4p_4}{T^4}+\left(\frac{121}{\epsilon^2}-\frac{51}{\epsilon}\right)\frac{p_2}{T^4}-\frac{22}{\epsilon}\frac{(4\pi)^2p_3}{T^4}\right]_{\epsilon^0}\,,
\end{align}
where the last two terms are the contributions of the coupling counterterm, which at two loops reads~\cite{Caswell:1974gg,Jones:1974mm}  
$\delta g_s^2=g_s^2[-11a_s/\epsilon+a_s^2(121/\epsilon^2-51/\epsilon)]$ with $a_s=g_s^2/(4\pi)^2$, and $p_2$ and $p_3$ must therefore be kept through $\mathcal{O}(\epsilon^2)$ and $\mathcal{O}(\epsilon)$, respectively.
This leads to our final result:
\begin{equation}
    q_c(0) = -154-3630.97-60.75=-3846\pm515\,,
\end{equation}
where the uncertainty combines in quadrature the allowances on the different sum-integrals, and the two errors on $\beta_G$. Comparing with $q_c=-3526(4)(55)(30)$, as obtained in \cite{Borsanyi:2012ve} by fitting the same $\mathcal{O}(g_s^6)$ formula to continuum-extrapolated lattice data for the trace anomaly in the range $10\,\Lambda_{\overline{\rm MS}}<T<1000\,\Lambda_{\overline{\rm MS}}$, we see that the two values differ merely by half a standard deviation.

\begin{figure}[t]
    \centering
    \includegraphics[width=\textwidth]{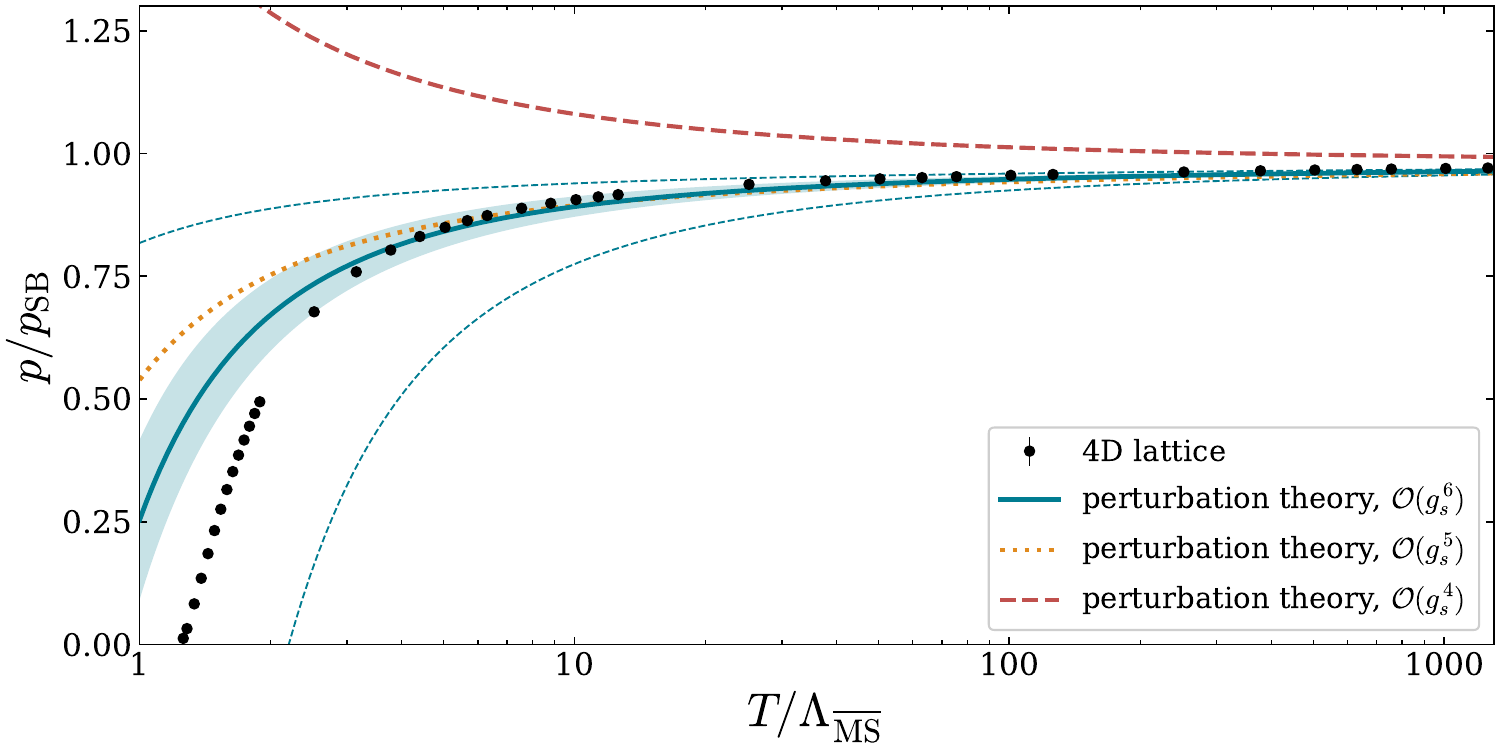}
    \caption{\it Pressure of the gluon plasma normalized to the Stefan--Boltzmann value, as a function of $T/\Lambda_{\overline{\rm MS}}$. The solid line is the strict expansion of Eq.~\eqref{eq:strict} with our $q_c(0)=-3846$, $\overline\mu=2\pi T$ and $g_s(\overline\mu)$ from the exact two-loop solution of the renormalization-group equation. The shaded band includes the effect of the uncertainty on the connected four-loop finite parts of Tab.~\ref{tab:sumintegrals}, and the dashed lines are the envelope obtained by varying $\pi T\leq\overline\mu\leq4\pi T$. The lattice data points are from \cite{Borsanyi:2012ve}, using their correspondence $T_c/\Lambda_{\overline{\rm MS}}\simeq1.26(7)$.}\label{fig:pressure}
\end{figure}

In Fig.~\ref{fig:pressure} we compare our result with the continuum-extrapolated lattice data of \cite{Borsanyi:2012ve}. The strict $\mathcal{O}(g_s^6)$ series with the computed $q_c(0)$ reproduces the lattice pressure within $3\%$ from $T\simeq2.5\,\Lambda_{\overline{\rm MS}}$ upwards, and within
$1.5\%$ from $T\simeq3\,\Lambda_{\overline{\rm MS}}$ to the end of the lattice range at $1000\,\Lambda_{\overline{\rm MS}}$. 

This is the first time that this comparison can be made without fitting the $\mathcal{O}(g_s^6)$ constant to the lattice data, and the agreement with the fitted value of \cite{Borsanyi:2012ve} discussed above is therefore a genuine test of the dimensionally-reduced weak-coupling framework at the first order at which its non-perturbative sector enters.

Finally, our calculation shows that the four-loop hard matching within QCD with quarks should also be within reach of the present technology. Fermionic sum-integrals can be handled using the method of Section~\ref{sec:numerical} by simply changing the Matsubara frequencies (and they are in fact IR-safer, as they have no zero modes). Important cross-checks would follow from a comparison with the already known $N_f^3$ contributions~\cite{Gynther:2009qf}, and selected lattice results for three~\cite{Bresciani:2025pio} and four~\cite{Borsanyi:2016ksw} families of quarks over a wide range of temperatures. We leave their exploration to future work.

\section*{Acknowledgments}
This work has been supported by the European Research Council under grant agreement n. 101230200. LG and JLM acknowledge support also from the FPU program under grants numbers FPU23/02026 and FPU23/02028, as well as from MICIU/AEI/10.13039/501100011033 and ERDF/EU (grants  PID2022-139466NB-C21, PID2022-139466NB-C22 and PID2025-170216NB-I00).

\section*{Note added}
In the final stages of this work, we became aware that the group in Helsinki was independently carrying out the same computation using entirely different methods~\cite{Kajantie:2026xxx}. We verified agreement for the single master sum-integral common to both calculations, as well as for the final result for $q_c(0)$ within uncertainties.

\section*{AI disclosure}
We declare that we have used generative AI tools (Claude and ChatGPT) for coding, as well as to discuss physics and mathematical questions, and to improve the writing of this manuscript. We reviewed, verified, and edited any AI output and we take full responsibility for the content of this work.

\appendix

\section{Pressure counterterms in the bosonic SM}
\label{app:counterterm}
We consider the superrenormalizable 3D EFT of the SM, in Euclidean space,
\begin{align}
    \mathcal{L}_\text{3D} &= \frac14 B_{ij}B_{ij}+\frac14 W^a_{ij}W^a_{ij}+\frac14 G^A_{ij}G^A_{ij}+(D_i\phi)^\dagger(D_i\phi)+\frac12(\partial_iB_0)^2+\frac12(D_iW_0^a)^2+\frac12(D_iG_0^A)^2\nonumber\\
    &+m_\phi^2\,\phi^\dagger\phi+\frac12m_{B_0}^2B_0^2+\frac12m_{W_0}^2W_0^aW_0^a+\frac12m_{G_0}^2G_0^AG_0^A\nonumber\\
    &+\lambda_{\phi^4}(\phi^\dagger\phi)^2+\lambda_{\phi^2B_0^2}\,\phi^\dagger\phi\,B_0^2+\lambda_{\phi^2W_0^2}\,\phi^\dagger\phi\,W_0^aW_0^a+\lambda_{\phi^2B_0W_0}\,B_0\,\phi^\dagger\sigma^a\phi\,W_0^a\,,
\end{align}
where $\phi$ is the (zeroth mode of the) Higgs doublet, with hypercharge $Y=1/2$; $B_0$, $W_0^a$ and $G_0^A$ are the temporal components of the $U(1)$, $SU(2)$ and $SU(3)$ gauge fields, which become scalars in the singlet and adjoint representations; $B_{ij}$, $W^a_{ij}$ and $G^A_{ij}$ are the field strengths of the spatial gauge fields; $D_i$ are the covariant derivatives with three-dimensional couplings $g_1$, $g_2$ and $g_s$\footnote{Note that this is not the same $g_s$ as the strong coupling of QCD, but its 3D equivalent, for which we use the same symbol.} and $SU(2)$ generators $\sigma^a/2$. We omit the self-interactions of the temporal scalars, as they do not enter the following counterterms. We use Feynman gauge ($\xi=1$) for the calculations below.

The two-loop counterterm for the pressure reads \cite{Kajantie:2002wa,Gynther:2005dj}

\begin{align}
    \delta^2 p_3 = \frac{1}{16 \pi^2\epsilon}\left(\frac{1}{4} g_1^2m_\phi^2 + \frac{3}{4} g_2^2 m_\phi^2 + \frac{3}{2} g_2^2 m_{W_0}^2 + 6g_s^2 m_{G_0}^2\right)\,.
\end{align}

The four-loop one, which includes four-loop diagrams as well as two-loop diagrams with mass counterterm insertions, that we compute here for the first time, reads

\begin{equation}\label{eq:4loopct}
    \delta^4 p_3 = \frac{1}{256\pi^4} \left[\frac{1}{\epsilon}(A_1 + \pi^2 A_2)+\frac{A_3}{\epsilon^2}\right]\,,
\end{equation}
with
\begin{align}
      A_1 &=-\frac{3}{32}g_1^6-\frac{51}{64}g_1^4g_2^2-\frac{69}{32}g_1^2g_2^4-\frac{2977}{64}g_2^6-387g_s^6-\frac{3}{8}(g_1^4+2g_1^2g_2^2+3g_2^4)\lambda_{\phi^4}\nonumber\\
      &-3(g_1^2+3g_2^2)\lambda_{\phi^4}^2-3g_2^4\lambda_{\phi^2W_0^2}-\frac{1}{2}(g_1^2+3g_2^2)\lambda_{\phi^2B_0^2}^2-\frac{3}{2}(g_1^2+11g_2^2)\lambda_{\phi^2W_0^2}^2 \nonumber\\
      &-\frac{3}{4}(g_1^2+7g_2^2)\lambda_{\phi^2B_0W_0}^2\,.
\end{align}

\begin{align}
    A_2 &=\frac{11}{1024}g_1^6+\frac{171}{2048}g_1^4g_2^2+\frac{233}{2048}g_1^2g_2^4+\frac{2827}{1024}g_2^6+\frac{729}{32}g_s^6+\frac{3}{64}(g_1^4+2g_1^2g_2^2+3g_2^4)\lambda_{\phi^4}\nonumber\\
    &+\frac{3}{32}(g_1^2+3g_2^2)\lambda_{\phi^4}^2+\frac{3}{8}g_2^4\lambda_{\phi^2W_0^2}+\frac{1}{64}(g_1^2+3g_2^2)\lambda_{\phi^2B_0^2}^2+\frac{3}{64}(g_1^2+11g_2^2)\lambda_{\phi^2W_0^2}^2\nonumber\\
    &+\frac{3}{128}(g_1^2+7g_2^2)\lambda_{\phi^2B_0W_0}^2+\frac{3}{2}\lambda_{\phi^4}^3+\frac{1}{6}\lambda_{\phi^2B_0^2}^3+\frac{1}{2}\lambda_{\phi^2W_0^2}^3+\frac{3}{8}\lambda_{\phi^4}\lambda_{\phi^2B_0^2}^2+\frac{9}{8}\lambda_{\phi^4}\lambda_{\phi^2W_0^2}^2\nonumber\\
    &+\frac{3}{16}\left[\lambda_{\phi^4}+2(\lambda_{\phi^2B_0^2}+\lambda_{\phi^2W_0^2})\right]\lambda_{\phi^2B_0W_0}^2\,,
\end{align}

\begin{align}
    A_3&=\frac{5}{512}g_1^6+\frac{33}{512}g_1^4g_2^2+\frac{15}{512}g_1^2g_2^4-\frac{21}{512}g_2^6-\frac{3}{32}(g_1^2+3g_2^2)^2\lambda_{\phi^4}+\frac{3}{8}(g_1^2+3g_2^2)\lambda_{\phi^4}^2\nonumber\\
    &-\frac{3}{4}g_2^2(g_1^2+3g_2^2)\lambda_{\phi^2W_0^2}+\frac{1}{16}(g_1^2+3g_2^2)\lambda_{\phi^2B_0^2}^2+\frac{3}{16}(g_1^2+11g_2^2)\lambda_{\phi^2W_0^2}^2 \nonumber\\
    &+\frac{3}{32}(g_1^2+7g_2^2)\lambda_{\phi^2B_0W_0}^2\,.
\end{align}

The expression in terms of 4D couplings can be trivially retrieved upon replacing the 3D couplings by their dimensional reduction matching~\cite{Kajantie:1995dw,Braaten:1995cm} (see \cite{Kajantie:2002wa} for the QCD case). Consequently, we omitted temporal-mode self-couplings in Eq.~\eqref{eq:4loopct} because they arise only at one-loop level.

\bibliographystyle{style}

\bibliography{references}

@article{Alloul:2013bka,
    author = "Alloul, Adam and Christensen, Neil D. and Degrande, C{\'e}line and Duhr, Claude and Fuks, Benjamin",
    title = "{FeynRules  2.0 - A complete toolbox for tree-level phenomenology}",
    eprint = "1310.1921",
    archivePrefix = "arXiv",
    primaryClass = "hep-ph",
    reportNumber = "CERN-PH-TH-2013-239, MCNET-13-14, IPPP-13-71, DCPT-13-142, PITT-PACC-1308",
    doi = "10.1016/j.cpc.2014.04.012",
    journal = "Comput. Phys. Commun.",
    volume = "185",
    pages = "2250--2300",
    year = "2014"
}

@article{Shtabovenko:2023idz,
    author = "Shtabovenko, Vladyslav and Mertig, Rolf and Orellana, Frederik",
    title = "{FeynCalc 10: Do multiloop integrals dream of computer codes?}",
    eprint = "2312.14089",
    archivePrefix = "arXiv",
    primaryClass = "hep-ph",
    reportNumber = "P3H-23-089, TTP23-056, SI-HEP-2023-27",
    doi = "10.1016/j.cpc.2024.109357",
    journal = "Comput. Phys. Commun.",
    volume = "306",
    pages = "109357",
    year = "2025"
}

@article{Hahn:2000kx,
    author = "Hahn, Thomas",
    title = "{Generating Feynman diagrams and amplitudes with FeynArts 3}",
    eprint = "hep-ph/0012260",
    archivePrefix = "arXiv",
    reportNumber = "KA-TP-23-2000",
    doi = "10.1016/S0010-4655(01)00290-9",
    journal = "Comput. Phys. Commun.",
    volume = "140",
    pages = "418--431",
    year = "2001"
}

@article{Kajantie:2003ax,
    author = "Kajantie, K. and Laine, M. and Rummukainen, K. and Schroder, Y.",
    title = "{Four loop vacuum energy density of the SU(N(c)) + adjoint Higgs theory}",
    eprint = "hep-ph/0304048",
    archivePrefix = "arXiv",
    reportNumber = "CERN-TH-2003-076, HIP-2003-16-TH, MIT-CTP-3354",
    doi = "10.1088/1126-6708/2003/04/036",
    journal = "JHEP",
    volume = "04",
    pages = "036",
    year = "2003"
}

@article{Pikelner:2017tgv,
    author = "Pikelner, Andrey",
    title = "{FMFT: Fully Massive Four-loop Tadpoles}",
    eprint = "1707.01710",
    archivePrefix = "arXiv",
    primaryClass = "hep-ph",
    reportNumber = "DESY-17-098",
    doi = "10.1016/j.cpc.2017.11.017",
    journal = "Comput. Phys. Commun.",
    volume = "224",
    pages = "282--287",
    year = "2018"
}

@article{Schroder:2025rka,
    author = {Schr{\"o}der, York and Stamou, Emmanuel and Steudtner, Tom and Uetrecht, Max},
    title = "{General Four-Loop Beta Function for Scalar-Fermion Theories in Three Dimensions}",
    eprint = "2512.02123",
    archivePrefix = "arXiv",
    primaryClass = "hep-th",
    doi = "10.21468/SciPostPhys.20.6.168",
    journal = "SciPost Phys.",
    volume = "20",
    pages = "168",
    year = "2026"
}

@article{Gil:2026cqz,
    author = "Gil, Luis and L{\'o}pez Miras, Javier and Moreno-S{\'a}nchez, Adri{\'a}n",
    title = "{SIRENA -- Sum-Integral REductioN Algorithm}",
    eprint = "2605.06775",
    archivePrefix = "arXiv",
    primaryClass = "hep-ph",
    month = "5",
    year = "2026"
}

@article{Braaten:1995jr,
    author = "Braaten, Eric and Nieto, Agustin",
    title = "{Free energy of QCD at high temperature}",
    eprint = "hep-ph/9510408",
    archivePrefix = "arXiv",
    reportNumber = "OHSTPY-HEP-T-95-020",
    doi = "10.1103/PhysRevD.53.3421",
    journal = "Phys. Rev. D",
    volume = "53",
    pages = "3421--3437",
    year = "1996"
}

@article{Kapusta:1979fh,
    author = "Kapusta, Joseph I.",
    title = "{Quantum Chromodynamics at High Temperature}",
    doi = "10.1016/0550-3213(79)90146-9",
    journal = "Nucl. Phys. B",
    volume = "148",
    pages = "461--498",
    year = "1979"
}

@article{Shuryak:1977ut,
    author = "Shuryak, Edward V.",
    title = "{Theory of Hadronic Plasma}",
    reportNumber = "IYF-77-34",
    journal = "Sov. Phys. JETP",
    volume = "47",
    pages = "212--219",
    year = "1978"
}

@article{Kajantie:2002wa,
    author = "Kajantie, K. and Laine, M. and Rummukainen, K. and Schroder, Y.",
    title = "{The Pressure of hot QCD up to g6 ln(1/g)}",
    eprint = "hep-ph/0211321",
    archivePrefix = "arXiv",
    reportNumber = "CERN-TH-2002-334, HIP-2002-62-TH, MIT-CTP-3325",
    doi = "10.1103/PhysRevD.67.105008",
    journal = "Phys. Rev. D",
    volume = "67",
    pages = "105008",
    year = "2003"
}

@article{Navarrete:2024ruu,
    author = {Navarrete, Pablo and Schr{\"o}der, York},
    title = "{The g$^{6}$ pressure of hot Yang-Mills theory: canonical form of the integrand}",
    eprint = "2408.15830",
    archivePrefix = "arXiv",
    primaryClass = "hep-ph",
    doi = "10.1007/JHEP11(2024)037",
    journal = "JHEP",
    volume = "11",
    pages = "037",
    year = "2024"
}

@article{Schwarz:1997gv,
    author = "Schwarz, Dominik J.",
    title = "{Evolution of gravitational waves through cosmological transitions}",
    eprint = "gr-qc/9709027",
    archivePrefix = "arXiv",
    reportNumber = "ETH-TH-97-26",
    doi = "10.1142/S0217732398002941",
    journal = "Mod. Phys. Lett. A",
    volume = "13",
    pages = "2771--2778",
    year = "1998"
}

@article{Chin:1978gj,
    author = "Chin, S. A.",
    title = "{Transition to Hot Quark Matter in Relativistic Heavy Ion Collision}",
    doi = "10.1016/0370-2693(78)90637-8",
    journal = "Phys. Lett. B",
    volume = "78",
    pages = "552--555",
    year = "1978"
}

@article{Toimela:1982hv,
    author = "Toimela, T.",
    title = "{The Next Term in the Thermodynamic Potential of {QCD}}",
    reportNumber = "HU-TFT-82-62",
    doi = "10.1016/0370-2693(83)91484-3",
    journal = "Phys. Lett. B",
    volume = "124",
    pages = "407--409",
    year = "1983"
}

@article{Arnold:1994ps,
    author = "Arnold, Peter Brockway and Zhai, Cheng-Xing",
    title = "{The Three loop free energy for pure gauge QCD}",
    eprint = "hep-ph/9408276",
    archivePrefix = "arXiv",
    reportNumber = "UW-PT-94-03",
    doi = "10.1103/PhysRevD.50.7603",
    journal = "Phys. Rev. D",
    volume = "50",
    pages = "7603--7623",
    year = "1994"
}

@article{Arnold:1994eb,
    author = "Arnold, Peter Brockway and Zhai, Cheng-xing",
    title = "{The Three loop free energy for high temperature QED and QCD with fermions}",
    eprint = "hep-ph/9410360",
    archivePrefix = "arXiv",
    reportNumber = "UW-PT-94-11",
    doi = "10.1103/PhysRevD.51.1906",
    journal = "Phys. Rev. D",
    volume = "51",
    pages = "1906--1918",
    year = "1995"
}

@article{DiRenzo:2006nh,
    author = "Di Renzo, F. and Laine, M. and Miccio, V. and Schroder, Y. and Torrero, C.",
    title = "{The Leading non-perturbative coefficient in the weak-coupling expansion of hot QCD pressure}",
    eprint = "hep-ph/0605042",
    archivePrefix = "arXiv",
    reportNumber = "BI-TP-2006-13",
    doi = "10.1088/1126-6708/2006/07/026",
    journal = "JHEP",
    volume = "07",
    pages = "026",
    year = "2006"
}

@article{Zhai:1995ac,
    author = "Zhai, Cheng-xing and Kastening, Boris M.",
    title = "{The Free energy of hot gauge theories with fermions through g**5}",
    eprint = "hep-ph/9507380",
    archivePrefix = "arXiv",
    reportNumber = "PURD-TH-95-03",
    doi = "10.1103/PhysRevD.52.7232",
    journal = "Phys. Rev. D",
    volume = "52",
    pages = "7232--7246",
    year = "1995"
}

@article{Gynther:2009qf,
    author = "Gynther, A. and Kurkela, A. and Vuorinen, A.",
    title = "{The N(f)**3 g**6 term in the pressure of hot QCD}",
    eprint = "0909.3521",
    archivePrefix = "arXiv",
    primaryClass = "hep-ph",
    reportNumber = "CERN-PH-TH-2009-172, TUW-09-14",
    doi = "10.1103/PhysRevD.80.096002",
    journal = "Phys. Rev. D",
    volume = "80",
    pages = "096002",
    year = "2009"
}

@article{Gross:1980br,
    author = "Gross, David J. and Pisarski, Robert D. and Yaffe, Laurence G.",
    title = "{QCD and Instantons at Finite Temperature}",
    reportNumber = "PRINT-80-0538 (PRINCETON)",
    doi = "10.1103/RevModPhys.53.43",
    journal = "Rev. Mod. Phys.",
    volume = "53",
    pages = "43",
    year = "1981"
}

@article{Ghisoiu:2015uza,
    author = "Ghisoiu, Ioan and Moller, Jan and Schroder, York",
    title = "{Debye screening mass of hot Yang-Mills theory to three-loop order}",
    eprint = "1509.08727",
    archivePrefix = "arXiv",
    primaryClass = "hep-ph",
    doi = "10.1007/JHEP11(2015)121",
    journal = "JHEP",
    volume = "11",
    pages = "121",
    year = "2015"
}

@article{Chala:2026wjp,
    author = "Chala, Mikael",
    title = "{Thermal masses in the Standard Model at three loops}",
    eprint = "2609.33448",
    archivePrefix = "arXiv",
    primaryClass = "hep-ph",
    month = "9",
    year = "2026"
}

@article{Ginsparg:1980ef,
    author = "Ginsparg, Paul H.",
    title = "{First Order and Second Order Phase Transitions in Gauge Theories at Finite Temperature}",
    reportNumber = "SACLAY-DPh-T 80/27",
    doi = "10.1016/0550-3213(80)90418-6",
    journal = "Nucl. Phys. B",
    volume = "170",
    pages = "388--408",
    year = "1980"
}

@article{Appelquist:1981vg,
    author = "Appelquist, Thomas and Pisarski, Robert D.",
    title = "{High-Temperature Yang-Mills Theories and Three-Dimensional Quantum Chromodynamics}",
    reportNumber = "Print-81-0020 (YALE), YTP-81-01, COO-3075-203",
    doi = "10.1103/PhysRevD.23.2305",
    journal = "Phys. Rev. D",
    volume = "23",
    pages = "2305",
    year = "1981"
}

@article{Braaten:1995cm,
    author = "Braaten, Eric and Nieto, Agustin",
    title = "{Effective field theory approach to high temperature thermodynamics}",
    eprint = "hep-ph/9501375",
    archivePrefix = "arXiv",
    reportNumber = "NUHEP-TH-95-2",
    doi = "10.1103/PhysRevD.51.6990",
    journal = "Phys. Rev. D",
    volume = "51",
    pages = "6990--7006",
    year = "1995"
}

@article{Kajantie:1995dw,
    author = "Kajantie, K. and Laine, M. and Rummukainen, K. and Shaposhnikov, Mikhail E.",
    title = "{Generic rules for high temperature dimensional reduction and their application to the standard model}",
    eprint = "hep-ph/9508379",
    archivePrefix = "arXiv",
    reportNumber = "CERN-TH-95-226, HU-TFT-95-50, IUHET-312",
    doi = "10.1016/0550-3213(95)00549-8",
    journal = "Nucl. Phys. B",
    volume = "458",
    pages = "90--136",
    year = "1996"
}

@article{Matsubara:1955ws,
    author = "Matsubara, Takeo",
    title = "{A New approach to quantum statistical mechanics}",
    doi = "10.1143/PTP.14.351",
    journal = "Prog. Theor. Phys.",
    volume = "14",
    pages = "351--378",
    year = "1955"
}

@article{Huovinen:2009yb,
    author = "Huovinen, Pasi and Petreczky, Pter",
    title = "{QCD Equation of State and Hadron Resonance Gas}",
    eprint = "0912.2541",
    archivePrefix = "arXiv",
    primaryClass = "hep-ph",
    doi = "10.1016/j.nuclphysa.2010.02.015",
    journal = "Nucl. Phys. A",
    volume = "837",
    pages = "26--53",
    year = "2010"
}

@article{Shen:2010uy,
    author = "Shen, Chun and Heinz, Ulrich and Huovinen, Pasi and Song, Huichao",
    title = "{Systematic parameter study of hadron spectra and elliptic flow from viscous hydrodynamic simulations of Au+Au collisions at $\sqrt{s_{NN}}=200$ GeV}",
    eprint = "1010.1856",
    archivePrefix = "arXiv",
    primaryClass = "nucl-th",
    doi = "10.1103/PhysRevC.82.054904",
    journal = "Phys. Rev. C",
    volume = "82",
    pages = "054904",
    year = "2010"
}

@article{Espinosa:2010hh,
    author = "Espinosa, Jose R. and Konstandin, Thomas and No, Jose M. and Servant, Geraldine",
    title = "{Energy Budget of Cosmological First-order Phase Transitions}",
    eprint = "1004.4187",
    archivePrefix = "arXiv",
    primaryClass = "hep-ph",
    reportNumber = "CERN-PH-TH-2010-027",
    doi = "10.1088/1475-7516/2010/06/028",
    journal = "JCAP",
    volume = "06",
    pages = "028",
    year = "2010"
}

@article{Asaka:2006rw,
    author = "Asaka, Takehiko and Laine, Mikko and Shaposhnikov, Mikhail",
    title = "{On the hadronic contribution to sterile neutrino production}",
    eprint = "hep-ph/0605209",
    archivePrefix = "arXiv",
    reportNumber = "TU-770, BI-TP-2006-17, CERN-PH-TH-2006-085",
    doi = "10.1088/1126-6708/2006/06/053",
    journal = "JHEP",
    volume = "06",
    pages = "053",
    year = "2006"
}

@article{Davydychev:2023jto,
    author = {Davydychev, Andrei I. and Navarrete, Pablo and Schr{\"o}der, York},
    title = "{Factorizing two-loop vacuum sum-integrals}",
    eprint = "2312.17367",
    archivePrefix = "arXiv",
    primaryClass = "hep-ph",
    doi = "10.1007/JHEP02(2024)104",
    journal = "JHEP",
    volume = "02",
    pages = "104",
    year = "2024"
}

@article{Harnett:2024nny,
    author = "Harnett, D. and Li, Siyuan and Steele, T. G.",
    title = "{Numerically computing finite temperature loop integrals using pySecDec}",
    eprint = "2401.05258",
    archivePrefix = "arXiv",
    primaryClass = "hep-ph",
    doi = "10.1140/epja/s10050-024-01323-5",
    journal = "Eur. Phys. J. A",
    volume = "60",
    number = "5",
    pages = "107",
    year = "2024"
}

@article{Niedermayer:2016ilf,
    author = "Niedermayer, Ferenc and Weisz, Peter",
    title = "{Massless sunset diagrams in finite asymmetric volumes}",
    eprint = "1602.03159",
    archivePrefix = "arXiv",
    primaryClass = "hep-lat",
    reportNumber = "MPP-2015-267",
    doi = "10.1007/JHEP06(2016)102",
    journal = "JHEP",
    volume = "06",
    pages = "102",
    year = "2016"
}

@article{Nishimura:2018wla,
    author = "Nishimura, Hiromichi and Korthals-Altes, Chris and Pisarski, Robert D. and Skokov, Vladimir",
    title = "{Thermal effective potential for the Polyakov loop to higher loop order}",
    eprint = "1811.11736",
    archivePrefix = "arXiv",
    primaryClass = "hep-th",
    doi = "10.22323/1.336.0155",
    journal = "PoS",
    volume = "Confinement2018",
    pages = "155",
    year = "2019"
}

@article{Cavalcanti:2021xtb,
    author = "Cavalcanti, Erich",
    title = "{Feynman amplitudes in periodically compactified spaces: Spin 0}",
    eprint = "2105.14425",
    archivePrefix = "arXiv",
    primaryClass = "hep-th",
    doi = "10.1103/PhysRevD.104.085019",
    journal = "Phys. Rev. D",
    volume = "104",
    number = "8",
    pages = "085019",
    year = "2021"
}

@article{Borsanyi:2012ve,
    author = "Borsanyi, Sz. and Endrodi, G. and Fodor, Z. and Katz, S. D. and Szabo, K. K.",
    title = "{Precision SU(3) lattice thermodynamics for a large temperature range}",
    eprint = "1204.6184",
    archivePrefix = "arXiv",
    primaryClass = "hep-lat",
    reportNumber = "WUB-12-09",
    doi = "10.1007/JHEP07(2012)056",
    journal = "JHEP",
    volume = "07",
    pages = "056",
    year = "2012"
}

@article{Laine:2005ai,
    author = "Laine, M. and Schroder, Y.",
    title = "{Two-loop QCD gauge coupling at high temperatures}",
    eprint = "hep-ph/0503061",
    archivePrefix = "arXiv",
    doi = "10.1088/1126-6708/2005/03/067",
    journal = "JHEP",
    volume = "03",
    pages = "067",
    year = "2005"
}

@article{Gynther:2007bw,
    author = "Gynther, A. and Laine, M. and Schroder, Y. and Torrero, C. and Vuorinen, A.",
    title = "{Four-loop pressure of massless O(N) scalar field theory}",
    eprint = "hep-ph/0703307",
    archivePrefix = "arXiv",
    doi = "10.1088/1126-6708/2007/04/094",
    journal = "JHEP",
    volume = "04",
    pages = "094",
    year = "2007"
}

@article{Schroder:2012hm,
    author = "Schroder, Y.",
    title = "{A fresh look on three-loop sum-integrals}",
    eprint = "1207.5666",
    archivePrefix = "arXiv",
    primaryClass = "hep-ph",
    doi = "10.1007/JHEP08(2012)095",
    journal = "JHEP",
    volume = "08",
    pages = "095",
    year = "2012"
}

@article{Ghisoiu:2012yk,
    author = "Ghisoiu, I. and Schroder, Y.",
    title = "{A new three-loop sum-integral of mass dimension two}",
    eprint = "1207.6214",
    archivePrefix = "arXiv",
    primaryClass = "hep-ph",
    doi = "10.1007/JHEP09(2012)016",
    journal = "JHEP",
    volume = "09",
    pages = "016",
    year = "2012"
}

@article{Ghisoiu:2012kn,
    author = "Ghisoiu, I. and Schroder, Y.",
    title = "{A new method for taming tensor sum-integrals}",
    eprint = "1208.0284",
    archivePrefix = "arXiv",
    primaryClass = "hep-ph",
    doi = "10.1007/JHEP11(2012)010",
    journal = "JHEP",
    volume = "11",
    pages = "010",
    year = "2012"
}

@article{Ellis:2016jkw,
    author = "Ellis, Joshua",
    title = "{TikZ-Feynman: Feynman diagrams with TikZ}",
    eprint = "1601.05437",
    archivePrefix = "arXiv",
    primaryClass = "hep-ph",
    doi = "10.1016/j.cpc.2016.08.019",
    journal = "Comput. Phys. Commun.",
    volume = "210",
    pages = "103--123",
    year = "2017"
}

@article{Laporta:2000dsw,
    author = "Laporta, S.",
    title = "{High-precision calculation of multiloop Feynman integrals by difference equations}",
    eprint = "hep-ph/0102033",
    archivePrefix = "arXiv",
    doi = "10.1142/S0217751X00002159",
    journal = "Int. J. Mod. Phys. A",
    volume = "15",
    pages = "5087--5159",
    year = "2000"
}

@article{Caswell:1974gg,
    author = "Caswell, William E.",
    title = "{Asymptotic Behavior of Nonabelian Gauge Theories to Two Loop Order}",
    reportNumber = "PRINT-74-1058 (PRINCETON)",
    doi = "10.1103/PhysRevLett.33.244",
    journal = "Phys. Rev. Lett.",
    volume = "33",
    pages = "244",
    year = "1974"
}

@article{Jones:1974mm,
    author = "Jones, D. R. T.",
    title = "{Two Loop Diagrams in Yang-Mills Theory}",
    reportNumber = "OXFORD-TP-10-74",
    doi = "10.1016/0550-3213(74)90093-5",
    journal = "Nucl. Phys. B",
    volume = "75",
    pages = "531",
    year = "1974"
}

@article{Gynther:2005dj,
    author = "Gynther, A. and Vepsalainen, M.",
    title = "{Pressure of the standard model at high temperatures}",
    eprint = "hep-ph/0510375",
    archivePrefix = "arXiv",
    reportNumber = "HIP-2005-46-TH",
    doi = "10.1088/1126-6708/2006/01/060",
    journal = "JHEP",
    volume = "01",
    pages = "060",
    year = "2006"
}

@article{Bresciani:2025pio,
    author = "Bresciani, Matteo and Brida, Mattia Dalla and Giusti, Leonardo and Pepe, Michele",
    title = "{Equation of State of QCD with Nf=3 flavours up to the electroweak scale}",
    eprint = "2509.26064",
    archivePrefix = "arXiv",
    primaryClass = "hep-lat",
    doi = "10.1016/j.jspc.2025.100252",
    journal = "J. Subatomic Part. Cosmol.",
    volume = "4",
    pages = "100252",
    year = "2025"
}

@article{Borsanyi:2016ksw,
    author = "Borsanyi, Sz. and others",
    title = "{Calculation of the axion mass based on high-temperature lattice quantum chromodynamics}",
    eprint = "1606.07494",
    archivePrefix = "arXiv",
    primaryClass = "hep-lat",
    reportNumber = "DESY-16-105",
    doi = "10.1038/nature20115",
    journal = "Nature",
    volume = "539",
    number = "7627",
    pages = "69--71",
    year = "2016"
}

@article{Kajantie:2026xxx,
    author = "Kajantie, Keijo and Navarrete, Pablo and Paatelainen, Risto and Sandbote, Leon and Schröder, York and Seppänen, Kaapo and Vuorinen, Aleksi",
    title = "Complete $g^6$ pressure of hot Yang-Mills theory",
    year = "2026",
    journal = "(to appear)"
}

\end{document}